\documentclass[twocolumn,showpacs,preprintnumbers,amsmath,amssymb,prl]{revtex4}
\usepackage{graphicx}% Include figure files
\usepackage{subfigure}
\usepackage{dcolumn}% Align table columns on decimal point
\usepackage{bm}% bold math
\usepackage{epstopdf}
\begin{document}

\title{Resistivity of the insulating phase approaching the 2D metal-insulator transition: the effect of spin polarization}
\author{Shiqi Li}
\affiliation{Department of Physics, City College of New York, CUNY,
New York, New York 10031, USA}
\author{M. P. Sarachik}
\affiliation{Department of Physics, City College of New York, CUNY,
New York, New York 10031, USA}

\date{\today}

\begin{abstract}

The resistivities of the dilute, strongly-interacting 2D electron systems in the insulating phase of a silicon MOSFET are the same for unpolarized electrons in the absence of magnetic field and for electrons that are fully spin polarized by the presence of an in-plane magnetic field. In both cases the resistivity obeys Efros-Shklovskii variable range hopping  $\rho(T) = \rho_0
\mbox{exp}[(T_{ES}/T)^{1/2}]$, with $T_{ES}$ and $1/\rho_0$  mapping onto each other if one applies a shift of the critical density $n_c$ reported earlier. With and withoug magnetic field, the parameters $T_{ES}$ and $1/\rho_0 = \sigma_0$ exhibit scaling consistent with critical behavior approaching a metal-insulator transition.

\end{abstract}

\pacs{71.30.+h, 72.20.Ee, 64.60.F-}

\maketitle

Two-dimensional (2D) electron systems realized in semiconductor
heterostructures and on the surface of doped semiconductor devices
such as silicon MOSFETs have been intensively studied for more than
half a century \cite{Ando1982}. Significant advances in fabrication
techniques in recent years have yielded samples with greatly
increased electron mobility thereby allowing access to lower
electron densities, a regime where the energy of interactions
between the electrons is dominant and substantially larger than
their kinetic energy. Rather than exhibiting a resistivity that
increases logarithmically toward infinity as the temperature is
reduced \cite{DolanDynes} as expected within the theory of
localization \cite{Abrahams1979}, the resistivity of these
strongly-interacting dilute 2D electron systems displays
metallic-like behavior at low temperatures and
insulating behavior as the electron density is reduced below a
material-dependent critical electron density $n_{\rm c}$ (see references
\cite{Abrahams2001,KravchenkoReports,Spivak2009} for reviews).  The apparent metal-insulator transition and metallic phase observed in high-mobility, strongly interacting 2D electron systems is widely regarded as one of the most important unresolved problems in condensed matter physics.

Exceptionally large magnetoresistances have been reported in
response to in-plane magnetic field in the vicinity
of the critical electron density, $n_c$. For electron densities $n_{\rm s} >
n_{\rm c}$ on the just-metallic side of the transition, increasing the parallel
magnetic field causes the resistivity to increase by several orders of magnitude at low temperatures and reach saturation at a density-dependent field $B_{\rm sat}$; for $B >  B_{\rm sat}$ the temperature dependence of the resistivity is
that of an insulator. Measurements have confirmed that the
saturation of the sample resistivity corresponds to full spin
polarization \cite{Okamoto1999, Vitkalov2000}.  Since the parallel
magnetic field does not couple to the orbital motion of electrons in
sufficiently thin 2D systems, this suggests an important role
for the electron spins.

The magnetoresistance has been thoroughly investigated on the metallic side of the transition, while considerably less information has been gathered in the insulating phase.  In order to better understand the effect of spin, we embarked on a
detailed comparison of the resistivity of the insulating state
arrived at by: (1) reducing the electron density below the critical
density $n_{\rm c}$ in zero field so that the electrons are unpolarized,
or (2) applying an in-plane magnetic field beyond the saturation
field $B_{\rm sat}$ where the metallic behavior is suppressed and
the spins are fully polarized.

It has been established in a number of experiments that the application of in-plane magnetic
field causes a shift of the critical density $n_c$
\cite{Simonian1997,Pudalov1997,Mertes1999,Jaroszynski2004,Shashkin2001}. As further detailed below, there have been conflicting reports on
the behavior of the resistivity in the insulating phase of
low-disorder 2D materials, with some claiming simply-activated
behavior and others claiming Efros-Shklovskii variable-range hopping
in the presence of a Coulomb gap due to electron interactions
\cite{Shklovskii1984}.

In this paper we report that both in zero field and in the presence
of an in-plane magnetic field sufficient to polarize all the
carriers, the resistivity obeys Efros-Shklovskii variable-range
hopping.  Moreover, we demonstrate that the unpolarized insulator
and the fully spin-polarized insulator map onto each other if one
simply shifts the critical density $n_c$. This implies that the
transport properties of the insulating state are the same in an
unpolarized and in a completely spin-polarized system.

Measurements were performed on silicon metal-oxide-semiconductor
field-effect transistors (MOSFETs) between 0.25 K and 2 K in an
Oxford Heliox He-3 refrigerator in the absence of magnetic field and
in a parallel field of 5 T. Similar to those used in
Ref.~\cite{Mokashi2012}, the high-mobility samples used in our
studies ($\mu_{\rm peak} = 3 \times 10^4 $ cm$^2$/Vs) were
fabricated in a Hall bar geometry of width 50 $\mu$m and distance
120 $\mu$m between the central potential probes; the oxide thickness
was 150 nm. Contact resistance was minimized by using a split-gate
geometry in which thin gaps are introduced in the gate metallization
so that a high electron density can be maintained near the contacts
independently of the value of the electron density in the main part
of the sample. Electron densities were controlled by applying a
positive dc gate voltage relative to the contacts.

\begin{figure}[h]
\centering
\includegraphics[width=0.5\textwidth]{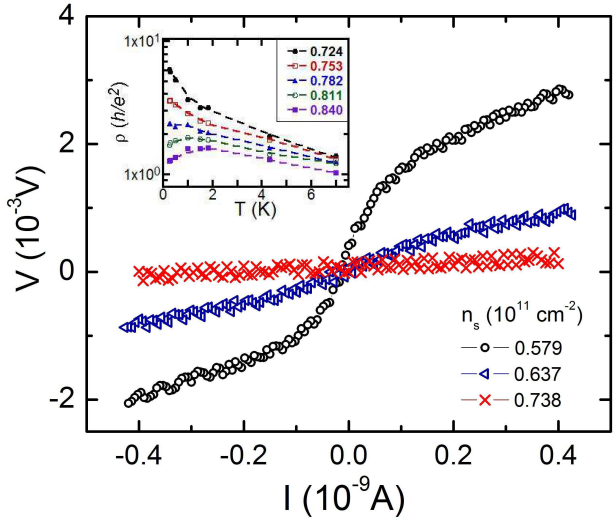}
\caption{\label{fig:nonlinearIV} Nonlinear current-voltage (I-V)
characteristics in zero field at several electron densities in the insulating regime; $T = 0.25$ K. As the electron
density approaches the critical density from the insulating side,
the nonlinearity of I-V gradually fades away \cite{Shashkin2001}. At
density $0.738 \times 10^{11}$ cm$^{-2}$ which is very close to the
critical density, the nonlinearity can only be seen when plotted on an
expanded scale. The inset shows the temperature dependence of
resistivity at a few electron densities near the critical density in
zero field.}
\end{figure}

As shown in Fig.~\ref{fig:nonlinearIV}, the voltage is a strongly
nonlinear function of the current: it exhibits linear (ohmic)
behavior at low currents and bends to a lower slope above a current
that depends on electron density and temperature. This is in
agreement with numerous past measurements and has variously been
attributed to non-ohmic strong electric field Efros-Shklovskii variable range hopping \cite{Marianer1992}, or the depinning of a Wigner Crystal or charge-density
wave \cite{Goldman1990, Kravchenko1991, D'Iorio1992, Pudalov1993,
Pudalov1994}. The resistivity plotted in the next few figures was
deduced from the low-current, linear portion of the curves measured
for each $n_{\rm s}$ and $T$ \cite{Mason1995}.

\begin{figure}[h]
\centering
\includegraphics[width=0.6\textwidth]{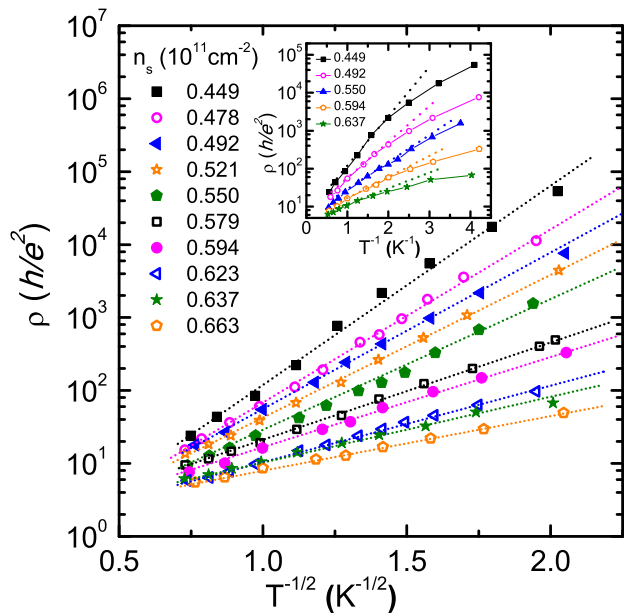}
\caption{\label{fig:rhoTB0} Resistivity vs $T^{-\frac{1}{2}}$ in
zero field for electron densities between $0.449$ and $0.663 \times
10^{11} $ cm$^{-2}$. For comparison the inset shows the data for
several densities plotted vs $T^{-1}$.}
\end{figure}

Figure \ref{fig:rhoTB0} shows the log of the resistivity measured in
zero field plotted against $T^{-\frac{1}{2}}$ for electron densities
$n_{\rm s} < n_{\rm c}$. For comparison, the inset shows the same
data plotted as a function of $T^{-1}$. Quite clearly, the data obey
the Efros-Shklovskii form of variable-range hopping in the
temperature range of the measurements (0.25 K to 2 K). Small
deviations at the lowest temperature may be due to poor thermal
contact of the electron system to the lattice (and thermometer).

\begin{figure}[h]
\centering
\includegraphics[width=0.57\textwidth]{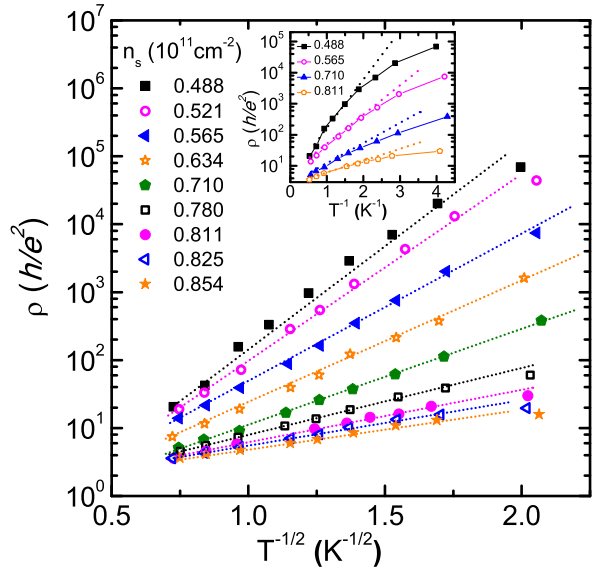}
\caption{\label{fig:rhoTB5} Resistivity vs $T^{-\frac{1}{2}}$ in 5 T
parallel magnetic field for electron densities between $0.488$ and
$0.654 \times 10^{11}$ cm$^{-2}$. The inset shows the data for
several densities plotted vs $T^{-1}$.}
\end{figure}

Similar data obtained in a parallel field $B_{||} = 5$ T are shown on
a semilog plot in Fig.~\ref{fig:rhoTB5}, as a function of
$T^{-\frac{1}{2}}$ in the main figure and as a function of $T^{-1}$
in the inset. Care was taken to check that $B_{||} = 5$ T is beyond
the saturation fields $B_{\rm sat}$ at the densities we used for the
measurement, so that all the spins in the sample are fully
polarized. The Efros-Shklovskii form of variable-range hopping,
$\rho(T)=\rho_0$ exp$[(T_{ES}/T)^{\frac{1}{2}}]$ provides an excellent
fit to the data in parallel magnetic field.

In a doped semiconductor at relatively low temperature where there is not enough (thermal)
energy to activate electrons across the mobility edge to conducting states in the impurity band, charge transport occurs via thermally
activated hopping between localized states, obeying the expression
\begin{equation}
\rho(T) = \rho_0 \mbox{exp}[(T_{ES}/T)^{\alpha}]
\end{equation}
Here the exponent $\alpha$ = 1 for nearest-neighbor hopping, $\alpha =
\frac{1}{3}$ for Mott variable-range hopping in two dimensions
\cite{Mott1969}, and $\alpha = \frac{1}{2}$ for Efros-Shklovskii variable-range hopping in the presence of a
Coulomb gap in the density of states associated with
electron-electron interactions \cite{Efros1975,Shklovskii1984}.

There have been a number of studies of the resistivity of silicon MOSFETs in the insulating phase.  The temperature dependence of the resistivity in zero
field was attributed to
Efros-Shklovskii variable range hopping with $\alpha = \frac{1}{2}$ down to $\approx 0.3$ K
by Mertes $et~al.$ \cite{Mertes1999} and Jarozynski $et~al.$
\cite{Jaroszynski2004} and over a broader range to lower
temperatures by Mason $et~al.$ \cite{Mason1995}. The zero-field
resistivity was found to obey the same form in a single
two-dimensional layer of $\delta$-doped GaAs/Al$_x$Ga$_{1-x}$As
\cite{Shlimak2000}. By contrast, Arrhenius-type activated behavior
($\alpha = 1$) was claimed for silicon in zero field by Shashkin
$et~al.$ \cite{Shashkin2001}.

Measurements in the presence of an in-plane magnetic field have
yielded various different results: Shashkin $et~al.$ fitted data for
silicon at intermediate temperatures and densities near the
transition to the activated form with $\alpha = 1$
\cite{Shashkin2001}; also in silicon, Mertes $et~al.$ applied a high
parallel magnetic field $B_{||} = 10.8$ T and found a larger
exponent $\alpha = 0.7$ \cite{Mertes2001}. For a single
two-dimensional layer of $\delta$-doped GaAs/Al$_x$Ga$_{(1-x)}$As in
parallel fields of 8 T and 6 T, Shlimak $et~al.$ reported an
exponent of 0.8 \cite{Shlimak2000}.

\begin{figure}[h]
\centering
\includegraphics[width=0.55\textwidth]{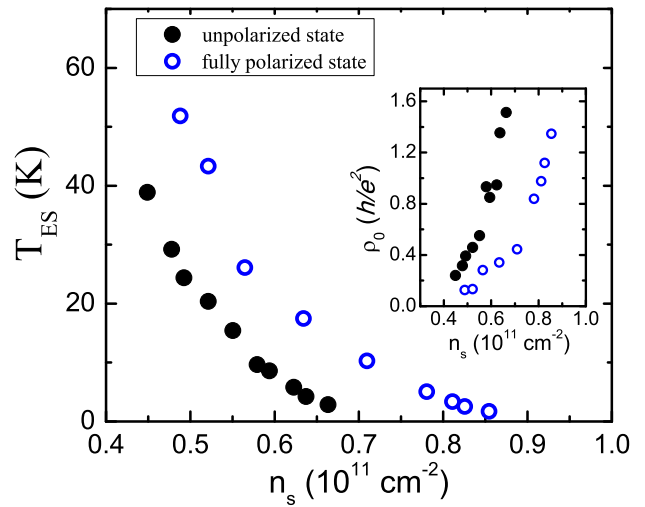}
\caption{\label{fig:TES} $T_{ES}(0)$ in zero field and $T_{ES}$(5T)$ =
T_{ES}(B_{\rm sat})$ in in-plane field of 5 T vs electron density.  The inset shows the parameter $\rho_0$ as a function of electron density.  Closed circles and open circles refer to data in zero field and $5$ T, respectively.}
\end{figure}

As shown in Fig.~\ref{fig:rhoTB0} and Fig.~\ref{fig:rhoTB5} above,
the data reported here in zero field and in 5 T parallel field are
both consistent with Efros-Shklovskii variable-range hopping,
$\rho(T)=\rho_0$exp$[(T_{ES}/T)^{\frac{1}{2}}]$ \cite{footnote}. The parameters $T_{ES}(0)$
in zero field and $T_{ES}$(5T) in 5 T parallel field are shown in
Fig.~\ref{fig:TES} as a function of electron density. $T_{ES}(0)$
and $T_{ES}$(5T) both decrease and gradually approach zero. While
$T_{ES}(0)$ extrapolates to zero at critical density $n_{\rm c}$ in
the absence of magnetic field, $T_{ES}$(5T) obtained in a 5 T
parallel field extrapolates to zero at a different electron density
which we designate as $n_{\rm c}$(5T) = $n_{\rm c}(B_{\rm sat})$. The inset to
Fig.~\ref{fig:TES} shows that the prefactor $\rho_0$ increases
sharply with increasing electron density.

The critical density $n_{\rm c}$ in zero field is generally determined as
the density at which the temperature derivative of the resistivity,
$d\rho/dT$, changes sign; an example of such a "separatrix" between
metallic and insulating behavior can be seen in the inset to
Fig.~\ref{fig:nonlinearIV}. The authors of Ref.~\cite{Shashkin2001}
have shown that in zero field the density where the nonlinearity of
current-voltage (I-V) characteristics vanishes yields the same
critical density. Using both protocols, the critical density in zero
field was estimated to be $n_{\rm c} = (0.782 \pm 0.014) \times 10^{11}$
cm$ ^{-2}$ for our sample.  We note that this is the critical density inferred from thermoelectric power measurements on the same sample \cite{Mokashi2012}.

An estimate of the critical density in a $5$ T
field can be obtained by determining the density at
which the nonlinearity in the I-V characteristic vanishes
\cite{Shashkin2001}. Using this procedure, an estimate is obtained
for the critical electron density at $B_{\rm sat}$ in our sample of
$n_{\rm c}(B_{\rm sat}) = (0.985 \pm 0.014) \times 10^{11} $ cm$^{-2}$.

In Efros and Shklovskii's theory of variable-range hopping
conduction for strongly interacting electrons, the parameter $T_{ES}
\propto \frac{e^2}{\epsilon \xi}$, where $\epsilon$ and $\xi$ are
the density-dependent dielectric constant and  localization length, respectively \cite{Shklovskii1984}. Moreover, $\epsilon(n_{\rm s}) \propto [n_{\rm c}/(n_{\rm c}-n_{\rm s})]^{\rm \zeta}$ and $\xi(n_{\rm s}) \propto
[n_{\rm c}/(n_{\rm c}-n_{\rm s})]^{\rm \nu}$ as electron density $n_{\rm s}$ approaches the critical density $n_{\rm c}$ from the insulating side, so that $T_{ES}$ obeys the critical form:

\begin{equation}
T_{ES} = A[(n_{\rm c}-n_{\rm s})/n_{\rm c}]^{\beta}
\end{equation}
where $\beta = (\zeta + \nu)$.

\begin{figure}[h]
\centering
\includegraphics[width=0.5\textwidth]{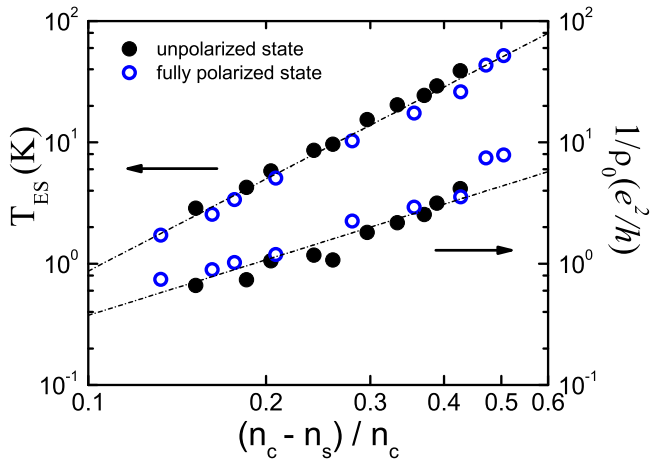}
\caption{\label{fig:compare} The parameters $T_{ES}$ and $1/\rho_0 = \sigma_0$ versus $[(n_{\rm c}-n_{\rm s})/n_{\rm c}]$ on a log-log scale.  The filled circles denote the measurements taken in zero field (for unpolarized electrons); the open circles denote measurements taken in a 5T in-plane magnetic field where the
 electrons are fully spin-polarized.  The value of $n_{\rm c}$ is deduced for each case as described in the text: $n_{\rm c}(B=0) = (0.782 \pm 0.014) \times 10^{11}$ cm$^{-2}$, $n_{\rm c}(B_{\rm sat}) = (0.985 \pm 0.014) \times 10^{11}$ cm$^{-2}$.}
\end{figure}

Using the values estimated for $n_{\rm c}(0)$ and $n_{\rm c}(B_{\rm sat})$, we now
plot $T_{ES}(0)$ in zero field (filled circles) and $T_{ES}(B_{\rm sat})$ in 5
T parallel field (triangles) as a function of $(n_{\rm c}-n_{\rm s})/n_{\rm c}$ and
$(n_{\rm c}(B_{sat})-n_{\rm s})/n_{\rm c}(B_{\rm sat})$, respectively, on a log-log scale
in Fig.~\ref{fig:compare}. $T_{ES}(0)$ in the unpolarized state and
$T_{ES}(B_{\rm sat})$ in the fully polarized state at 5 T lie on the same
curve. Relative to the appropriate critical density for each case,
the Efros-Shklovskii parameter $T_{ES} \propto \frac{1}{\epsilon \xi}$
approaches the metallic phase with the same critical exponent, $T_{ES}
= A[(n_{\rm c}-n_{\rm s})/n_{\rm c}]^{\beta}$ with $\beta = 2.5 \pm 0.1$.  Interestingly, the prefactor $1/\rho_0 = \sigma_0$ is also consistent with critical behavior.

Although our data and many past experiments are consistent with the occurrence of a transition, it is noteworthy and puzzling that the value we've obtained for the critical exponent $\beta$ is substantially different from  the smaller exponents in the vicinity of $1.6$ reported in a number of earlier studies \cite{exp1.6}.  Our analysis is based on data obtained quite far from the transition and entails two parameters, $\rho_0$ and $T_{ES}$, while earlier analyses using a single parameter $T_0$ were based on data obtained near and on both sides of the presumed critical point.  A careful examination shows that the portion of our data that is near the transition is consistent with one-parameter scaling and a smaller exponent of $\beta = 1.7$, while the portion of the published data that is further from the transition is can be scaled using two parameters, $T_{ES}$ and $\rho_0$ with $\beta$ much closer to our value of $2.5$ \cite{footnote2}.  Thus, there is no discrepancy between data sets, and the difference may be due to relative closeness to the critical regime.

As suggested in Ref. \cite{Mokashi2012}, there may indeed be two quantum critical points in play: $n_c$ driven by disorder, and a disorder-independent universal interaction-driven critical point $n_i$. They are different in principle, but so close to each other in the low-disorder samples of our studies that they have not been separately identified experimentally. The scaling of $T_{ES}$ and $\rho_0$ is determined using data obtained in the insulating phase only, and it is by no means clear which (if either) critical point it refers to. This raises the possibility that there may be two closely spaced transitions, one driven by disorder and one driven by interactions.  In the range of densities just below the transition where the temperature derivative of the resistivity, $d\rho/dT$, is negative indicating insulating behavior, the resistivity is a very weak function of temperature so that it cannot be reliably fit to the Efros-Shklovskii form.  It is thus possible that there is an intermediate phase in this region that gives rise to different exponents entering and leaving this phase as the density is increased.  Additional careful studies to lower temperatures of samples with yet lower disorder in this range of densities would be of great interest.

To summarize, we have measured the resistivity of the dilute,
strongly-interacting 2D electron system in a silicon MOSFET in the
insulating phase arrived at by (1) reducing the electron density in
the absence of magnetic field, which results in no spin polarization
(2) applying a 5 Tesla in-plane field which results in complete spin
polarization.  For both cases, the resistivity obeys
Efros-Shklovskii variable-range hopping with parameters $T_{ES}$ and
$\rho_0$ that are consistent with critical behavior approaching a
metal-insulator transition.  The sole effect of spin polarization is
a simple shift of the critical density. The fact that the transport
properties of the insulating state are the same in two systems with
different spin configurations should be very interesting and worth
further theoretical attention.

Useful comments were provided by Steve Kivelson, Boris Spivak,
Vladimir Dobrosavljevic and Dragana Popovic.  We thank Boris
Shklovskii, Dietrich Belitz and Sergey Kravchenko for numerious
discussions, valuable insights and for their critical reading of
this manuscript.  This work was supported by the National Science
Foundation grant DMR-1309008 and the Binational Science Foundation
Grant 2012210.


\begin{thebibliography}{99}
\bibitem{Ando1982} For example, see T. Ando, A. B. Fowler and F. Stern, Rev. Mod. Phys. {\bf 54}, 437 (1982).
\bibitem{DolanDynes} G. J. Dolan and D. D. Osheroff, Phys. Rev. Lett. {\bf 43}, 721
(1979); D. J. Bishop, D. C. Tsui, and R. C. Dynes, Phys. Rev. Lett.
{\bf 44}, 1153 (1980); M. J. Uren, R. A. Davies, and M. Pepper, J. Phys. C
{\bf 13}, L985 (1980).
\bibitem{Abrahams1979} E. Abrahams, P. W. Anderson, D. C. Licciardello, and
T. V. Ramakrishnan, Phys. Rev. Lett. {\bf42}, 673 (1979).
\bibitem{Abrahams2001} E. Abrahams, S. V. Kravchenko, and M. P. Sarachik,
Rev. Mod. Phys. 73, 251 (2001).
\bibitem{KravchenkoReports} S. V. Kravchenko and M. P. Sarachik,
Rep. Prog. Phys. {\bf67}, 1 (2004).
\bibitem{Spivak2009} B. Spivak, S. V. Kravchenko S. A. Kivelson, and X. P.
A. Gao, Rev. Mod. Phys. {\bf82}, 1743 (2009).
\bibitem{Okamoto1999} Tohru Okamoto, Kunio Hosoya, Shinji Kawaji,
and Atsuo Yagi, Phys. Rev. Lett. {\bf82}, 3875 (1999).
\bibitem{Vitkalov2000} S. A. Vitkalov, Hairong Zheng, K. M. Mertes, M. P. Sarachik, and T. M.
Klapwijk, Phys. Rev. Lett. {\bf85}, 2164 (2000); S. A. Vitkalov, M.
P. Sarachik and T. M. Klapwijk, Phys. Rev. B {\bf 64}, 073101
(2001).
\bibitem{Simonian1997} D. Simonian, S. V. Kravchenko, M. P. Sarachik, and V. M. Pudalov, Phys. Rev. Lett. {\bf79}, 2304 (1997)
\bibitem{Pudalov1997} V. M. Pudalov, G. Brunthaler, A. Prinz, and G. Bauer,
JETP Lett. {\bf65}, 932 (1997).
\bibitem{Mertes1999} K. M. Mertes, D. Simonian, M. P. Sarachik, S.
V. Kravchenko, and T. M. Klapwijk, Phys. Rev. B {\bf60}, R5093 (1999).
\bibitem{Shashkin2001} A. A. Shashkin, S. V. Kravchenko, and T. M. Klapwijk, Phys. Rev. Lett. {\bf87}, 266402 (2001).
\bibitem{Jaroszynski2004} J. Jaroszynski, Dragana Popovic, and T. M. Klapwijk, Phys. Rev. Lett. {\bf92}, 226403 (2004).
\bibitem{Shklovskii1984} B. I. Shklovskii, A. L. Efros, Electronic Properties of Doped Semiconductors, Solid State Series Vol. 45
(Springer-Verlag Berlin Heidelberg, 1984).
\bibitem{Mokashi2012} A. Mokashi, S. Li, Bo Wen, S. V. Kravchenko, A. A. Shashkin, V. T. Dolgopolov, M. P. Sarachik, Phys. Rev. Lett. {\bf109}, 096405 (2012).
\bibitem{Marianer1992} S. Marianer and B. I. Shklovskii, Phys. Rev. B {\bf 46}, 13100 (1992).
\bibitem{Goldman1990} V. J. Goldman, M. Santos, M. Shayegan, and J. E. Cunningham, Phys. Rev. Lett. {\bf65}, 2189 (1990).
\bibitem{Kravchenko1991} S. V. Kravchenko, V. M. Pudalov, J. Campbell, and M. D'Iorio, JETP Lett.{\bf54}, 532 (1991).
\bibitem{D'Iorio1992} M. D'Iorio, V. M. Pudalov, and S. G. Semenchinsky, Phys. Rev. B {\bf46}, 15992 (1992).
\bibitem{Pudalov1993} V. M. Pudalov, M. D'Iorio, S. V. Kravchenko, and J. W. Campbell, Phys. Rev. Lett. {\bf70}, 1866 (1993).
\bibitem{Pudalov1994} V. M. Pudalov and S. T. Chui, Phys. Rev. B {\bf49}, 14062 (1994).
\bibitem{Mason1995} W. Mason, S. V. Kravchenko, G. E. Bowker, and J. E. Furneaux, Phys. Rev. B {\bf52}, 7857 (1995).
\bibitem{Mott1969} N. F. Mott, Philosophical Magazine, {\bf19}, 835 (1969).
\bibitem{Efros1975} A. L. Efros and B. I. Shklovskii, J. Phys. C, {\bf L49}, 8 (1975).
\bibitem{Shlimak2000} I. Shlimak, S. I. Khondaker, M. Pepper, and D. A. Ritchie, Phys. Rev. B {\bf61}, 07253 (2000).
\bibitem{Mertes2001} K. M. Mertes, Hairong Zheng, S. A. Vitkalov, M. P. Sarachik, and T. M. Klapwijk, Phys. Rev. B {\bf63}, 041101(R) (2001).
\bibitem{footnote} Acceptable fits to our data can be obtained to the expression $\rho = A T^x\mbox{exp}[(T_{ES}/T)^p$ using a temperature-dependent prefactor. This yields a strongly negative value of $x$ for $p=1$, a small positive value of x consistent with $0$ for $p=1/2$ and $x \approx 2$ for $p=1/3$. Although little is known about the range of acceptable values for $x$, these numbers suggest that $p=1/2$ and $x \approx 0$ (a temperature-independent prefactor) is a reasonable choice. Data over a broader range of temperature could restrict the range for acceptable fits.
\bibitem{exp1.6} S. V. Kravchenko, Whitney E. Mason, G. E. Bowker, J. E. Furneaux, V. M. Pudalov, and M. D'Iorio, Phys. Rev. B {\bf 51}, 7038 (995); S. V. Kravchenko, D. Simonian, M. P. Sarachik, Whitney Mason, and J. E. Furneaux, Phys. Rev. Lett. {\bf 77}, 4938 (1996); Dragana Popovic, A. B. Fowler, and S. Washburn, Phys. Rev. Lett. {\bf 79}, 1543 (1997); X. G. Feng, Dragana Popovic, and S. Washburn, Phys. Rev. Lett. {\bf 83}, 368 (1999).
\bibitem{footnote2}  The value $2.5$ is very close to the critical exponent $2.3$ expected for a classical percolation transition. However, the disorder potential in Si is short-range so the possibility of a percolation transition can be safely ruled out.



\end{thebibliography}
\end{document}